\documentclass[aps,pra,reprint,superscriptaddress,showpacs,floatfix,longbibliography]{revtex4-1}

\usepackage{graphicx}
\usepackage{amssymb} 
\usepackage{dcolumn}
\usepackage{bm}
\usepackage[mathlines]{lineno}
\usepackage[colorlinks,linkcolor=blue,anchorcolor=blue,citecolor=blue,urlcolor=blue]{hyperref}
\usepackage{multirow}
\usepackage{color, soul}

\makeatletter
\renewcommand*{\@fnsymbol}[1]{\ensuremath{\ifcase#1\or \dagger\or *\or \ddagger\or
\mathsection\or \mathparagraph\or \|\or **\or \dagger\dagger \or
\ddagger\ddagger \else\@ctrerr\fi}} \makeatother

\usepackage{color}%

\begin{document}
\title{%
\hfill{\normalsize\rm Phys. Rev. Applied {\bf 10} (2018)}\\
Two-dimensional Mechanical Metamaterials with Unusual Poisson
Ratio Behavior}


\author{Zhibin Gao}
\affiliation{Physics and Astronomy Department,
             Michigan State University,
             East Lansing, Michigan 48824, USA}
\affiliation{Center for Phononics and Thermal Energy Science,
             China-EU Joint Center for Nanophononics,
             Shanghai Key Laboratory of Special Artificial
             Microstructure Materials and Technology,
             School of Physics, Sciences and Engineering,
             Tongji University, Shanghai 200092, China}

\author{Dan Liu}
\affiliation{Physics and Astronomy Department,
             Michigan State University,
             East Lansing, Michigan 48824, USA}

\author{David Tom\'anek} %
\email[E-mail: ]{tomanek@msu.edu}
\affiliation{Physics and Astronomy Department,
             Michigan State University,
             East Lansing, Michigan 48824, USA}

\date{\today }
\begin{abstract}
We design two-dimensional (2D) mechanical metamaterials that may
be deformed substantially at little or no energy cost. Examples of
such deformable structures are assemblies of rigid isosceles
triangles hinged in their corners on the macro-scale, or
polymerized phenanthrene molecules forming porous graphene on the
nano-scale. In these and in a large class of related structures,
the Poisson ratio $\nu$ diverges for particular strain values.
$\nu$ also changes its magnitude and sign, and displays a %
shape 
memory effect.
\end{abstract}

\pacs{
61.46.-w,    
64.70.Nd,    
73.22.-f,    
81.05.Zx     
 }


\maketitle




\section{Introduction}

There is growing interest in mechanical metamaterials, man-made
structures with counter-intuitive mechanical
properties~\cite{Bertoldi2017}. Unlike in ordinary uniform
materials, deformations in such metamaterials derive from the
geometry of the assembly rather than the elastic properties of the
components. This behavior is scale independent, covering
structures from the macro- to the nanoscale. Most attention in
this respect seems to be drawn by the Poisson ratio
$\nu$~\cite{greaves2011}, the negative ratio of lateral to applied
strain. Ordinary materials with typical values $0<{\nu}<0.5$
contract laterally when stretched, %
with unusually large values reported for cellular
materials~\cite{gibson1982mechanics}.
Auxetic metamaterials with ${\nu}<0$, on the other hand, expand in
both directions when
stretched~\cite{%
{gibson1982mechanics},
{Lakes87},%
{Lakes93},%
{baughman1993crystalline},%
{alderson1999triumph}} %
leading to advanced functionalities~\cite{%
{Mitschke2011},%
{gao2017novel}}. %
Auxetic systems with macroscopic components have been utilized for
shock absorption in automobiles~\cite{RR-patent}, in
high-performance clothing~\cite{%
{papadopoulou2017auxetic},%
{Chen17},%
{Nike-patent}}, in bioprostheses~\cite{scarpa2008auxetic} and
stents~\cite{Liner-patent} in medicine, and for strain
amplification~\cite{baughman1998negative}. Auxetic 2D mechanical
metamaterials with nanostructured components,
%
some of which have been described previously%
~\cite{%
{Shan15},{Julian93},{Wojciechowski89},{grima2000self}%
}, %
may find their use when precise micromanipulation of 2D structures
including bilayer graphene is
required~\cite{cao2018unconventional}.

Here we report the design of 2D mechanical metamaterials that may
be deformed substantially at little or no energy cost. Unlike
origami- and kirigami-inspired metamaterials, which derive their
functionality from folding a 2D material into the third
dimension~\cite{{schenk2013},{yasuda15},{Rafsanjani17},%
{grima2015tailoring}}, the structures we describe are confined to
a plane during deformation. Such confinement may be achieved by a
strong attraction to a planar substrate or in a sandwich geometry.
Specifically, we consider %
infinite %
assemblies of rigid isosceles triangles hinged in their corners on
the macro-scale~\cite{GuestHutchinson03} and polymerized
phenanthrene molecules forming `porous graphene' on the
nano-scale. In these and in a large class of related structures,
consisting of connected and near-rigid isosceles triangles, the
Poisson ratio $\nu$ diverges at particular strain values. $\nu$
also changes its magnitude and sign, and displays a `shape memory'
effect in a specific range of deformations, meaning that this
quantity depends on previously applied strain. Our corresponding
results are scale invariant.

\begin{figure*}
\includegraphics[width=1.8\columnwidth]{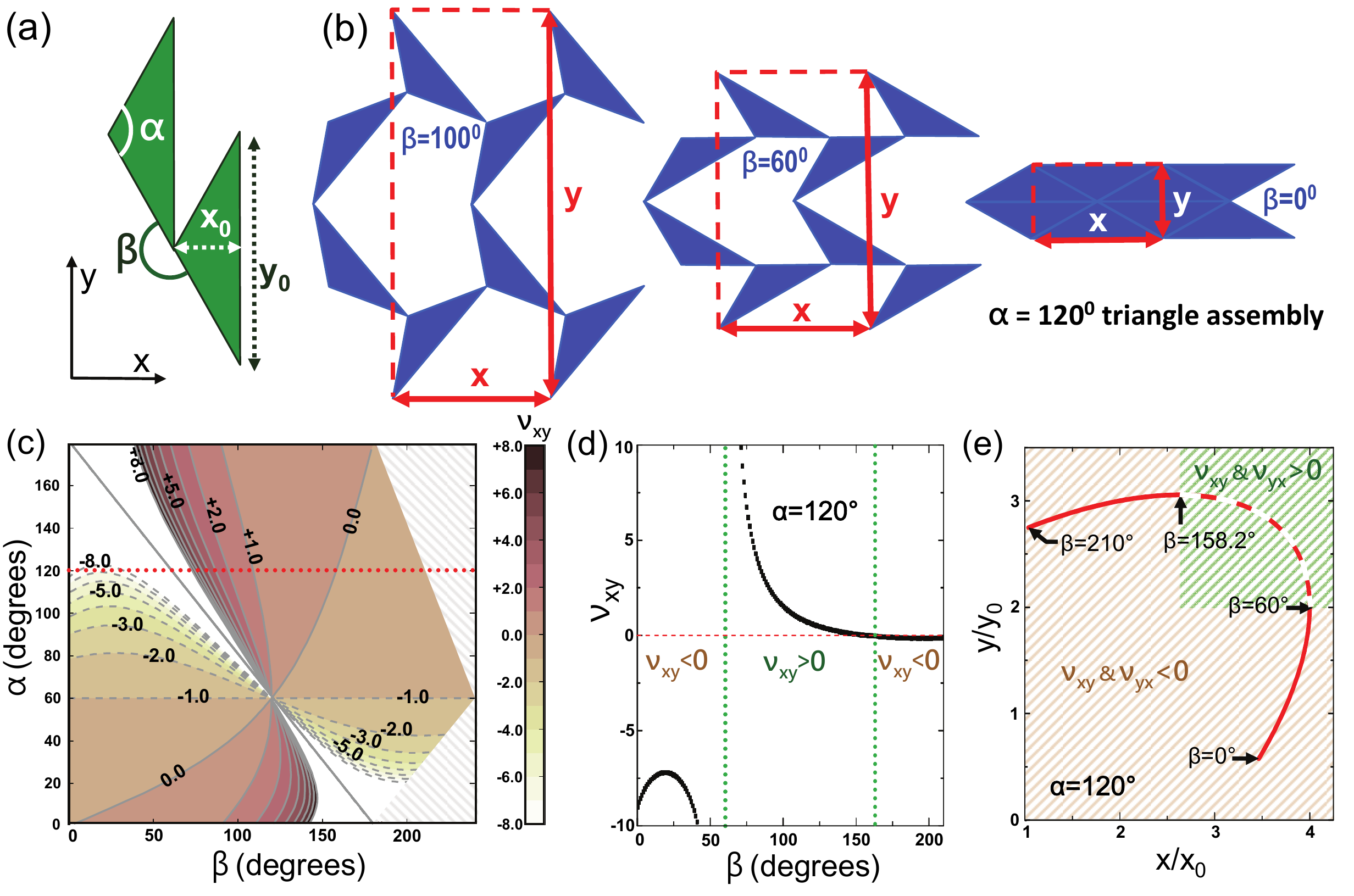}
\caption{Deformations in a 2D assembly of rigid
isosceles triangles. %
(a) Adjacent triangles with opening angle $\alpha$ and mutual
orientation defined by the closing angle $\beta$, hinged
tip-to-corner, forming the primitive unit cell. The triangle
height $x_0$ and the length $y_0$ of its base define %
the horizontal and vertical length scales. %
(b) Snap shots of the $\alpha=120^\circ$ triangle assembly for
different values of $\beta$. The conventional rectangular unit
cell is twice the size of the primitive unit cell. %
(c) Contour plot of the Poisson ratio $\nu_{xy}=-(dy/y)/(dx/x)$ as
a function of $\alpha$ and $\beta$. The dotted red line
highlights behavior of the ${\alpha}=120^\circ$ triangle assembly. %
(d) Poisson ratio $\nu_{xy}$ as a
function of $\beta$ in the $\alpha=120^\circ$ system. %
(e) Changes in the scaled width $x/x_0$ and height $y/y_0$ of the
conventional unit cell for $\alpha=120^\circ$ caused by changing
the angle $\beta$. %
\label{fig1}}
\end{figure*}

\begin{figure*}
\includegraphics[width=1.8\columnwidth]{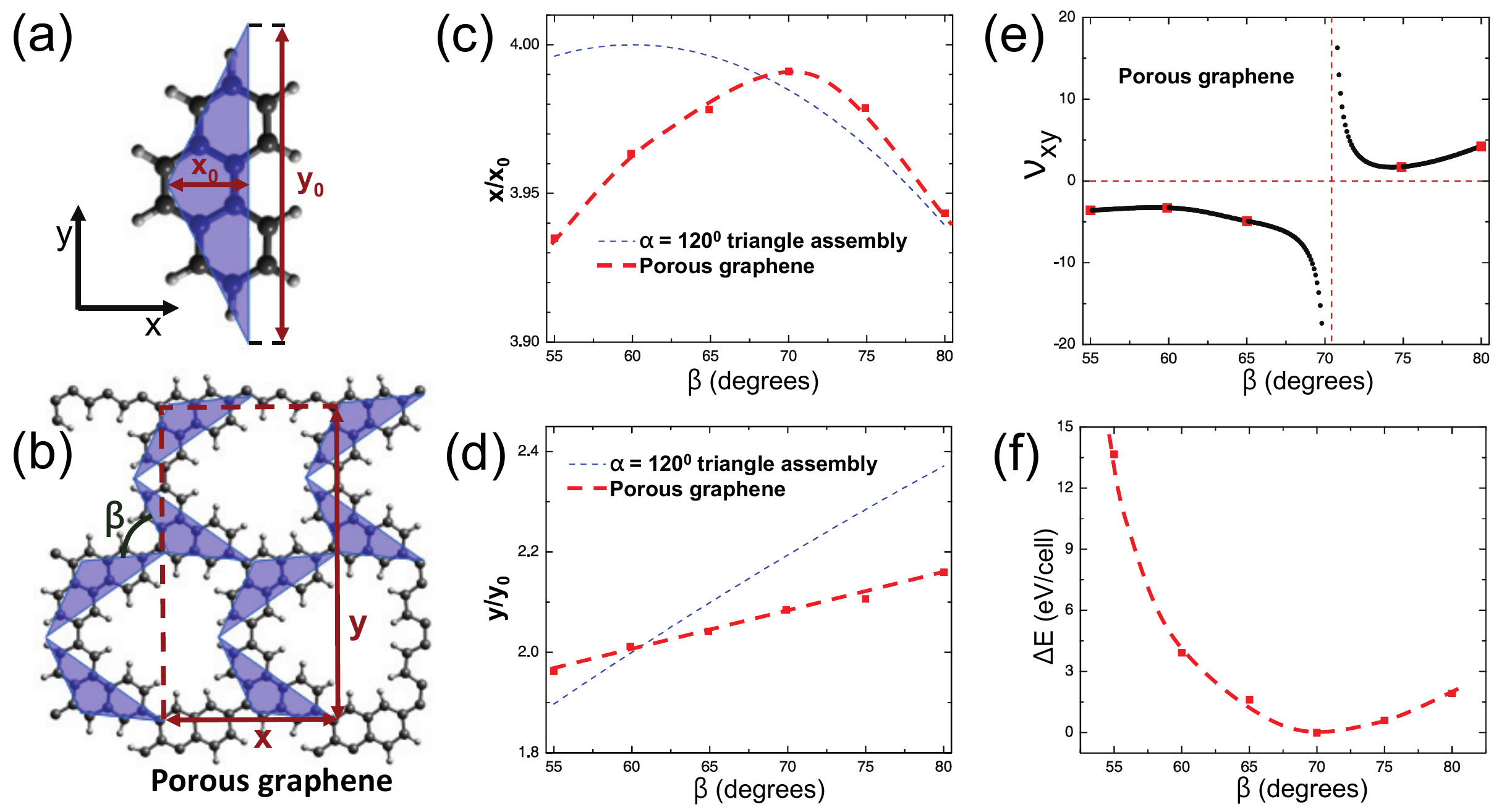}
\caption{Deformations in porous graphene, a
phenanthrene-based 2D mechanical metamaterial. %
(a) Structure of the C$_{14}$H$_{10}$ phenanthrene molecule and
its relation to an isosceles ${\alpha}=120^\circ$
triangle of Fig.~\protect\ref{fig1}. %
(b) Equilibrium structure of 2D porous graphene consisting of
polymerized phenanthrene molecules with ${\beta}=70^\circ$.
Saturating hydrogen atoms are shown by the lighter and smaller
spheres. %
Changes in the scaled width $x/x_0$ (c) and height $y/y_0$ (d) of
the conventional unit cell in the triangle assembly and porous
graphene as a function of the closing angle $\beta$. %
(e) Poisson ratio $\nu_{xy}$ in porous graphene as a function
of $\beta$. %
(f) Strain energy in the C$_{56}$H$_{28}$ conventional unit cell
as a function of $\beta$. %
The dashed and dotted lines connecting data points for porous
graphene in (c)-(f) are guides to the eye. %
\label{fig2}}
\end{figure*}

\section{Computational Approach}

We have studied the electronic and structural properties as well
as the deformation energy of polyphenanthrene dubbed `porous
graphene' using {\em ab initio} density functional theory (DFT) as
implemented in the \textsc{VASP}
code~\cite{{VASP1},{VASP2},{VASP3}}. %
We represented this 2D structure
by imposing periodic boundary conditions in all directions and
separating individual
layers by a vacuum region of $20$~{\AA}. We used
projector-augmented-wave (PAW)
pseudopotentials~\cite{{PAW1},{PAW2}} and the
Perdew-Burke-Ernzerhof (PBE)~\cite{PBE} exchange-correlation
functional. The Brillouin zone of the conventional unit cell of
the 2D structure has been sampled by an $5{\times}3{\times}1$
$k$-point grid~\cite{Monkhorst-Pack76}. We used $500$~eV as the
electronic kinetic energy cutoff for the plane-wave basis and a
total energy difference between subsequent self-consistency
iterations below $10^{-4}$~eV as the criterion for reaching
self-consistency. All geometries have been optimized using the
conjugate-gradient method~\cite{CGmethod}, until none of the
residual Hellmann-Feynman forces exceeded $10^{-2}$~eV/{\AA}.

\section{Results}

\subsection{Constructing a 2D mechanical metamaterial}

Figure~\ref{fig1} depicts the macro-scale 2D mechanical
metamaterial we consider, namely an %
infinite assembly of rigid isosceles triangles hinged in the
corners and described using periodic boundary conditions. %
There are two identical triangles with different orientation in
the primitive unit cell of the lattice, as seen in
Fig.~\ref{fig1}(a). The conventional unit cell, shown in
Fig.~\ref{fig1}(b), is rectangular and twice the size of the
primitive unit cell. The deformation behavior of such constrained
lattices of polygons including rectangles~\cite{grima2005auxetic}
and connected bars,
%
some of which display a Poisson ratio that changes sign,
value, and even diverges, %
%
has been described and classified
earlier~\cite{{GuestHutchinson03},{milton2013complete}}. In our
system, structural changes are regulated by the only independent
variable, the angle $\beta$. The full range of $\beta$ is
$0~{\leq}~{\beta}~{\leq}~{\alpha}+180^\circ$ for
${\alpha}{\le}60^\circ$ and
$0~{\leq}~{\beta}~{\leq}~270^\circ-{\alpha}/2$ for
${\alpha}{\ge}60^\circ$. Since there is no energy involved when
changing $\beta$, the structure maintains its geometry
after deformation. Snap shots of the triangle assembly and the
conventional unit cell at different values of $\beta$, shown in
Fig.~\ref{fig1}(b), illustrate the unusual flexibility of the
system. The movie of the continuous shape change is provided in
Video~\ref{Video1} in the Appendix.

For a system of triangles aligned with the Cartesian coordinate
system as shown in Fig.~\ref{fig1}(a), we can determine the strain
in the $y$-direction in response to strain applied along the
$x$-direction. The negative ratio of these strains is the Poisson
ratio $\nu_{xy}$, which is given by
%
\begin{eqnarray}
\nu_{xy}\!&=&\! -\frac{dy/y}{dx/x} \nonumber \\%
          &=& \frac{
         \cos(\frac{\alpha}{2})\sin(\frac{\beta}{2})
       -3\sin(\frac{\alpha}{2})\cos(\frac{\beta}{2}) %
       }%
       {%
       \cos(\frac{\alpha}{2})\cos(\frac{\beta}{2}) +
       3\sin(\frac{\alpha}{2})\sin(\frac{\beta}{2}) %
       }%
       \!\tan\!\left(\!\frac{\alpha\!+\!\beta}{2}\right)\!.
       \label{Eq1}%
\end{eqnarray}
Dependence of $\nu_{xy}$ on $\alpha$ and $\beta$ is presented as a
contour plot in Fig.~\ref{fig1}(c). Several aspects of this result
are noteworthy when inspecting the behavior of $\nu_{xy}({\beta})$
for a constant value of the opening angle $\alpha$. With the
exception of ${\alpha}=60^\circ$ describing equilateral
triangles~\cite{{grima2000zeolites},{Sun12369}}, $\nu_{xy}$
changes magnitude and sign with changing $\beta$. Presence of the
tangent function in Eq.~(\ref{Eq1}) causes $\nu_{xy}$ to diverge
to $\pm\infty$ for ${\beta}_{crit}(\nu_{xy})=180^\circ-\alpha$,
with ${\beta}_{crit}(\nu_{xy})=60^\circ$ for ${\alpha}=120^\circ$.
For ${\alpha}>60^\circ$, $\nu_{xy}$ changes sign twice across the
full range of $\beta$ values, as shown in Fig.~\ref{fig1}(d) for
${\alpha}=120^\circ$. The condition for the divergence of
$\nu_{yx}=1/\nu_{xy}$, describing strain in the $x$-direction in
response to strain applied in the $y$-direction, is
$\tan({\beta}_{crit}(\nu_{yx})/2)=3\tan({\alpha}/2)$. For
${\alpha}=120^\circ$, $\nu_{yx}$ will diverge at
${\beta}_{crit}(\nu_{yx})=158.2^\circ$.

Maybe the most unexpected aspect of our result is the %
`shape memory' effect displayed by both $\nu_{xy}$ and $\nu_{yx}$
if the angle $\beta$ becomes a hidden variable in the system. To
explain what we mean, we first %
inspect the $(x(\beta)/x_0, y(\beta)/y_0)$ trajectory given by
\begin{eqnarray}
\frac{x}{x_0}&=& %
  2 \left[ \tan\left(\frac{\alpha}{2}\right)
           \cos\left(\frac{\beta}{2}\right) +
           \sin\left(\frac{\beta}{2}\right) \right]\,,
  \label{Eq2} \\
\frac{y}{y_0}&=& %
  3 \sin\left(\frac{\beta}{2}\right) +%
    \cot\left(\frac{\alpha}{2}\right)
    \cos\left(\frac{\beta}{2}\right)\,.
  \label{Eq3}
\end{eqnarray}
The $(x(\beta)/x_0, y(\beta)/y_0)$ trajectory, describing the
changing shape of the unit cell, is shown for ${\alpha}=120^\circ$
in Fig.~\ref{fig1}(e), and for other values of $\alpha$ in
Fig.~\ref{fig4} in the Appendix section.
The sign of the
slope of the trajectory, opposite to the sign of $\nu_{xy}$ and
$\nu_{yx}$, changes twice as the structure unfolds with increasing
$\beta$. Regions of positive and negative $\nu_{xy}$ and
$\nu_{yx}$, delimited by the above-mentioned critical values
${\beta}_{crit}(\nu_{xy})$ for $\nu_{xy}$ and
${\beta}_{crit}(\nu_{yx})$ for $\nu_{yx}$, are distinguished
graphically in Fig.~\ref{fig1}(e). For any $x$ in the range
$3.46<x/x_0<4.00$, there are two different values of $y$
associated with different values of $\beta$ and different signs of
$\nu_{xy}$. Similarly, for any $y$ in the range $2.75<y/y_0<3.06$,
there are two different solutions for $x$ associated with
different values of $\beta$ and different signs of $\nu_{yx}$.

Let us now consider a macroscopic piece of `material' consisting
of hinged triangles, which are so small that their mutual
orientation cannot be made out. %
With no information about the deformation history, the %
material may exhibit either a positive or a negative Poisson
ratio. The {\em only} way to change the material so that it would
exhibit a definite positive or negative sign of the Poisson ratio
is to subject it to a sequence of deformations. Assume that this
material is first stretched to its maximum along a given direction
such as $x$. Subsequent stretching along a direction normal to the
first will result in a positive, subsequent compression in a
negative Poisson ratio. We may say that the system retains a
memory of previous deformations.

What happens microscopically can be clearly followed in
Fig.~\ref{fig1}(e). Even though the value of $\beta$ is hidden, we
know that it becomes $60^\circ$ for maximum stretch along $x$ and
$158.2^\circ$ for maximum stretch along $y$. Subsequent
deformation normal to the first direction then dictates the sign
of $\nu$. %
This behavior derives from the nonlinearity in the system and, in
some aspect, parallels the behavior of shape memory alloys.


\subsection{Porous graphene as a 2D mechanical metamaterial}

Whereas macroscopic triangular assemblies with various values of
$\alpha$ will %
find their use in particular applications, we turn our interest to
2D nanostructures that can be formed by coordination chemistry and
macromolecular assembly. Microstructures including colloidal
Kagom{\'e}
lattices~\cite{{chen2011directed},{Hiroshi16},{Hiroshi17}} and
graphitic nanostructures~\cite{{treier2011surface},{Moreno18}}
including polyphenylene~\cite{Porgra09}, sometimes dubbed
nanoporous graphene, have been synthesized, but do not display a
negative Poisson ratio. In the following, we focus on
polyphenanthrene, a 2D structure of phenanthrene molecules shown
in Fig.~\ref{fig2}(a). There is a strong similarity between this
molecule and ${\alpha}=120^\circ$ triangles depicted in
Fig.~\ref{fig1}. In particular, 2D assemblies of structures in
Figs.~\ref{fig1}(a) and \ref{fig2}(a) display strong similarities
in their Poisson ratio behavior discussed below.


The calculated equilibrium structure of 2D porous graphene formed
of polymerized phenanthrene molecules with the optimum angle
${\beta}=70^\circ$, shown in Fig.~\ref{fig2}(b), illustrates the
relationship between this structure and the ${\alpha}=120^\circ$
triangle assembly. The unusual flexibility of polyphenanthrene is
owed to the connection of phenanthrene molecules by strong C-C
$\sigma$ bonds, which are also responsible for the strength and
flexibility of polyethylene. %
Our DFT calculations indicate only small structural distortions of
the phenanthrene molecules, which nevertheless break their initial
mirror symmetry. %

In Fig.~\ref{fig2}(c) we compare changes in the scaled width
$x/x_0$ of the conventional unit cell as a function of the closing
angle $\beta$ for the assembly of triangles and for porous
graphene. The corresponding changes in the scaled height $y/y_0$
are shown in Fig.~\ref{fig2}(d) in the same range of $\beta$
values. Interestingly, $x({\beta})/x_0$ reaches its maximum at
${\beta}_{crit}(\nu_{xy})$ for both systems, whereas
$y({\beta})/y_0$ increases monotonically with increasing $\beta$.
According to the definition of the Poisson ratio
$\nu_{xy}=-(dy/y)/(dx/x)$, $\nu_{xy}$ diverges at
${\beta}_{crit}(\nu_{xy})=60^\circ$ in the triangular assembly, as
seen in Fig.~\ref{fig1}(d). Similarly, $\nu_{xy}$ diverges at
${\beta}_{crit}(\nu_{xy})=70^\circ$ in porous graphene, as shown
in Fig.~\ref{fig2}(e). The slope of $x({\beta})/x_0$ changes sign
at ${\beta}_{crit}$, resulting in $\nu_{xy}<0$ for
${\beta}<{\beta}_{crit}(\nu_{xy})$ and $\nu_{xy}>0$ for
${\beta}>{\beta}_{crit}(\nu_{xy})$ in both systems.

\begin{figure}[t]
\includegraphics[width=0.7\columnwidth]{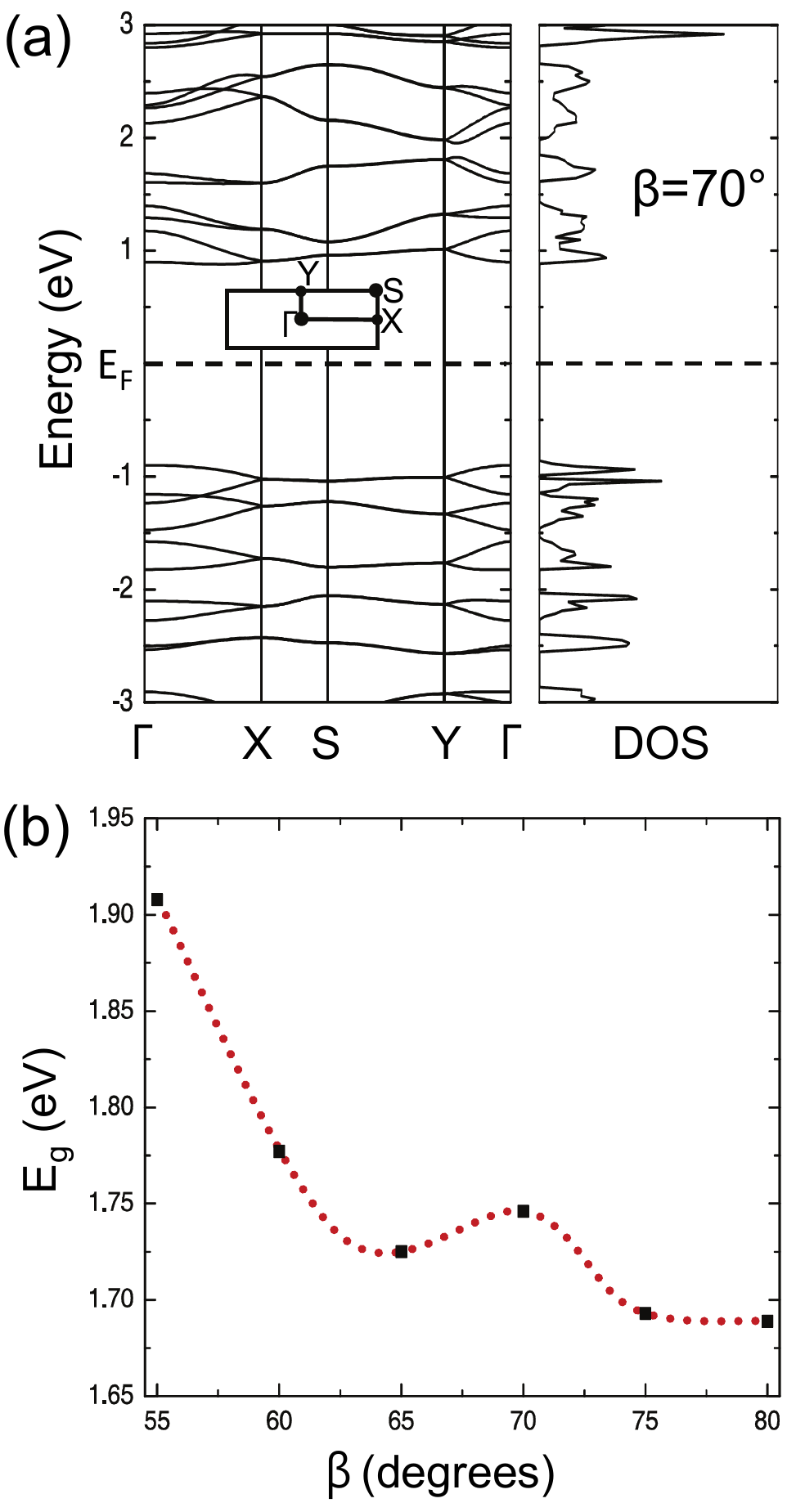}
\caption{ Electronic structure of porous graphene, a
phenanthrene-based 2D mechanical metamaterial, based on
DFT-PBE calculations. %
(a) Band structure of the equilibrium structure with
${\beta}=70^\circ$ obtained using the rectangular C$_{56}$H$_{28}$
unit cell. High-symmetry points in the rectangular Brillouin zone
are shown in the inset. %
(b) Fundamental band gap $E_g$ as a function of the angle $\beta$. %
\label{fig3}}
\end{figure}

The energy investment ${\Delta}E$ associated with deforming the
polyphenanthrene structure is shown in Fig.~\ref{fig2}(f). Our
results were obtained by optimizing the structure for selected
values of the angle $\beta$ that defines the relative orientation
of the two inequivalent phenanthrene molecules in the unit cell.
With ${\beta}{\approx}70^\circ$ representing the structural
optimum, we found that changing $\beta$ by ${\pm}10^\circ$
required ${\Delta}E<3$~eV per unit cell, corresponding to an
energy investment of only ${\approx}50$~meV per C atom, about 1\%
of the bond breaking energy. Thus, the polyphenanthrene structure
is rather soft %
and represents a valid counterpart to the isoenergetic model
system of Fig.~\ref{fig1}. %

Phenanthrene is a tricyclic organic molecule with a $3.36$~eV wide
DFT-PBE gap between the lowest unoccupied molecular orbital (LUMO)
and the highest occupied molecular orbital (HOMO). When
polymerized to the 2D polyphenanthrene structure depicted in
Fig.~\ref{fig2}(b), the HOMO broadens to the valence and the LUMO
to the conduction band. This is seen in Fig.~\ref{fig3}(a), which
depicts the band structure and the density of states of the
optimum geometry of polyphenanthrene with ${\beta}=70^\circ$, with
the Brillouin zone shown in the inset. Our DFT-PBE results
indicate that the fundamental band gap $E_g$ is reduced from the
molecular value to 1.75~eV in the equilibrium structure of the
layer, but still does not vanish for $55^\circ<{\beta}<80^\circ$.
The gap is near-direct due to the flatness of bands, and decreases
from $1.9$~eV at ${\beta}=55^\circ$ to $1.7$~eV at
${\beta}=80^\circ$. We should remember that Kohn-Sham eigenvalues
in all DFT calculations including ours do not correctly represent
the electronic structure and typically underestimate the band
gaps.

The decrease of $E_g$ and its dependence on $\beta$ upon
polymerization is caused by the presence of covalent C-C bonds
that connect individual phenanthrene molecules elastically and
electronically. Unfolding of the polyphenanthrene structure with
increasing angle $\beta$ rotates individual phenanthrene molecules
and modifies the bonding at the connection between adjacent
monomers, causing the the electronic structure to depend on
$\beta$. The range of deformations in polyphenanthrene is smaller
than in triangular assemblies due to the steric hindrance caused
by hydrogen termination. In absence of planar confinement,
phenanthrene molecules rotate out-of-plane at large tensile strain
values not considered here.


\section{Discussion}

Elastic response of materials is commonly described by elastic
constants constituting the elastic matrix, which describe
stress-strain relationships and thus contain energy in their
dimension. The Poisson ratio is fundamentally different. It is a
dimensionless quantity that describes deformations induced by
strain, independent of the energy cost.
According to its definition in Eq.~(\ref{Eq1}), it depends on the
choice of the coordinate system. The trace of the strain matrix,
however, which describes the fractional change of the area induced
by the mechanism, is independent of the choice of coordinates and
could couple naturally to external fields such as pressure. %

We believe that changes in pore size caused by the deformation of
the 2D unit cell may find their use in
tunable sieving in a
layered system~\cite{{Bernhard18},{filter2018}},
including application in desalination membranes. %
2D mechanical metamaterials may also find unusual applications in
micro-manipulation. In particular, a 2D layer in partial contact
with an in-plane junction of 2D metamaterials with different
values of $\nu$, including ${\nu}>0$ and ${\nu}<0$, may experience
a torque normal to the plane when in-plane strain is applied at
the junction of the 2D systems. Also the observation of
strain-related electronic structure changes in polyphenanthrene
opens new possibilities. Since polyphenanthrene and a wide range
of porous graphene structures can be viewed as a system of
covalently connected quantum dots, in-layer strain may be used to
tune the coupling between such quantum dots and thus change the
electronic structure of the system.

\section{Summary and Conclusions}

In summary, we have designed 2D mechanical metamaterials that may
be deformed substantially at little or no energy cost. Unlike
origami- and kirigami-based mechanical metamaterials that derive
their functionality from folding a 2D material to the third
dimension, the structures we design are confined to a plane during
deformation. In reality, such confinement may be achieved by a
strong attraction to a planar substrate or in a sandwich geometry.
On the macro-scale, the structures we describe are assemblies of
rigid isosceles triangles hinged in their corners. Their nanoscale
counterpart are molecules such as phenanthrene that may be
polymerized using coordination chemistry or macromolecular
assembly to form specific geometries with a porous graphene
structure. In these and in a large class of related structures,
consisting of connected and near-rigid isosceles triangles
confined to a plane, the Poisson ratio $\nu$ diverges for
particular strain values. $\nu$ also changes its magnitude and
sign, depending on the applied uniaxial strain, and
displays a %
shape 
memory effect with respect to the deformation history.

\section{Appendix}
\setcounter{equation}{0}
\renewcommand{\theequation}{A\arabic{equation}}

\subsection{Deformation behavior in 2D isosceles triangle assemblies}

\begin{video}[h]
\includegraphics[width=0.4\columnwidth]{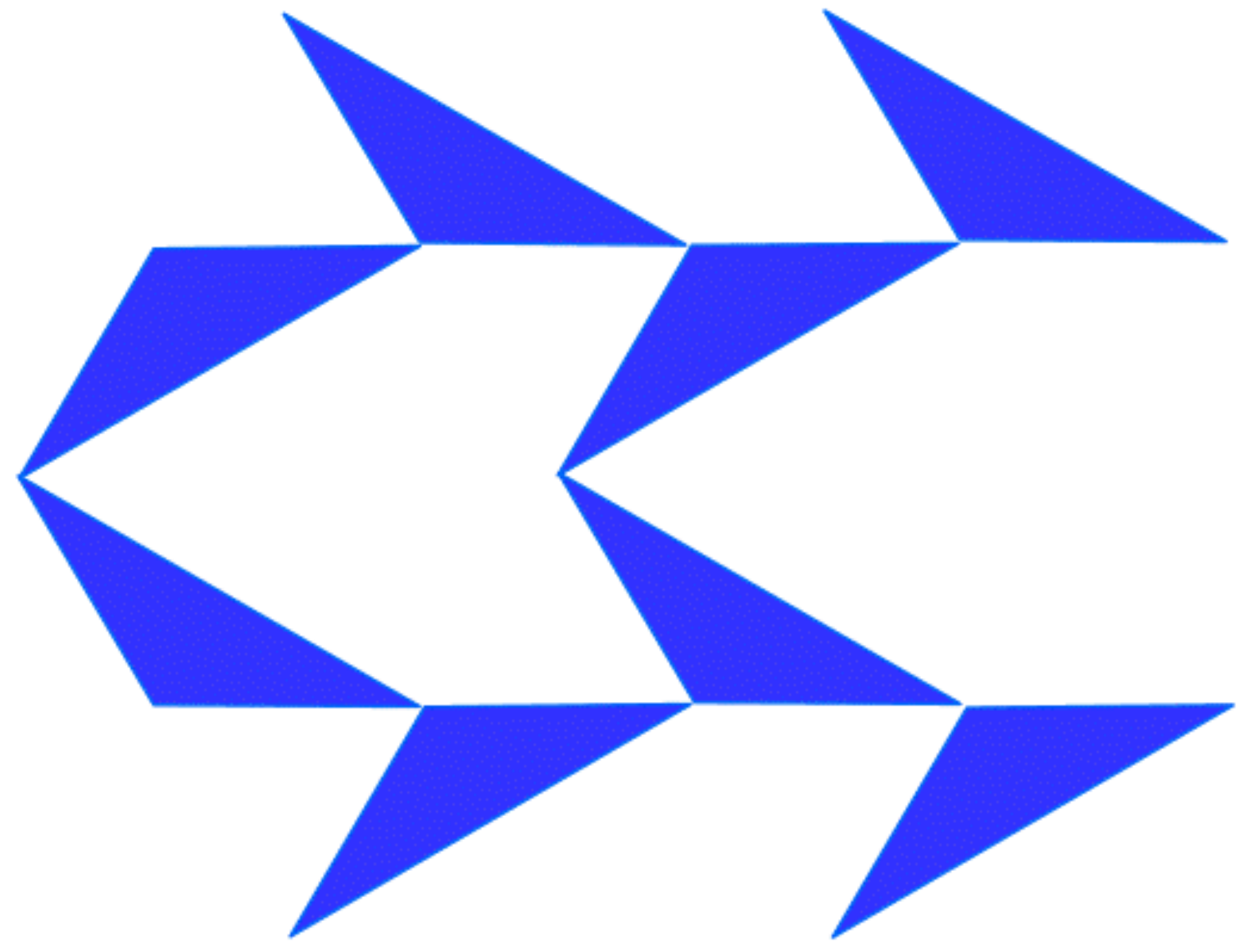}
\setfloatlink{Video1.mp4} %
\caption{Unfolding of a 2D assembly of ${\alpha}=120^\circ$
isosceles triangles with changing angle $\beta$. %
\label{Video1}}
\end{video}

As discussed earlier,
for a given value $y$ of
the unit cell height in a 2D assembly of isosceles triangles with
${\alpha}>60^\circ$, we can find two different values $x$ of the
unit cell width, with the two structures displaying opposite signs
of $\nu$. Similarly, we can find two different values $y$ for a
given value of $x$, with the two structures displaying opposite
signs of $\nu$. This unusual behavior results from the presence of
a hidden variable, the relative triangle orientation $\beta$, and
causes $\nu$ to depend not only on the overall sample shape, but
also the history of the system. The unfolding of an assembly of
triangles with ${\alpha}=120^\circ$ and its history dependence has
been characterized by the $x-y$ trajectory in Fig.~\ref{fig1}(e)
in the range of accessible $\beta$ angles. The unfolding process
of the ${\alpha}=120^\circ$ triangle assembly is depicted in
Video~\ref{Video1}.

\begin{figure}[b]
\includegraphics[width=0.6\columnwidth]{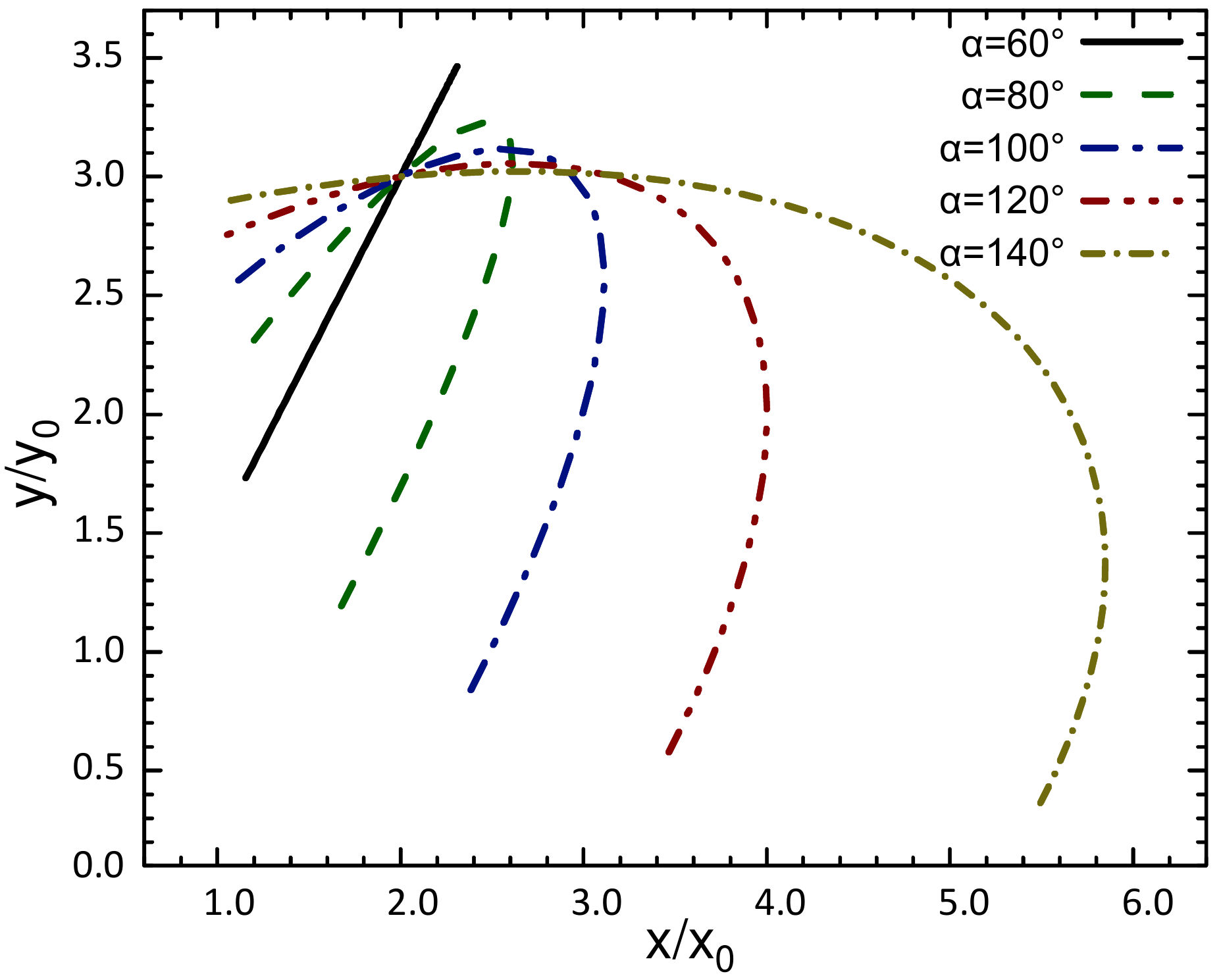}
\caption{ Changes in the scaled width $x/x_0$ and height $y/y_0$
of the conventional unit cell for different values of the opening
angle $\alpha$ as a function of the closing angle $\beta$. The
relevant quantities are defined in Fig.~1.
\label{fig4}}
\end{figure}

$x-y$ trajectories for several values of $\alpha$ are shown in
Fig.~\ref{fig4}. The particular shape of these $x-y$ trajectories
indicates that also for opening angles other than
${\alpha}=120^\circ$ discussed above, the value and sign of $\nu$
may depend on sample history. Only in the specific case of
equilateral triangles with ${\alpha}=60^\circ$, discussed in the
following, the $y-x$ trajectory in Fig.~\ref{fig4} is linear and
$\nu$ is history independent.

\subsection{Deformations in a 2D assembly of rigid equilateral triangles}

\begin{figure}[h]
\includegraphics[width=1.0\columnwidth]{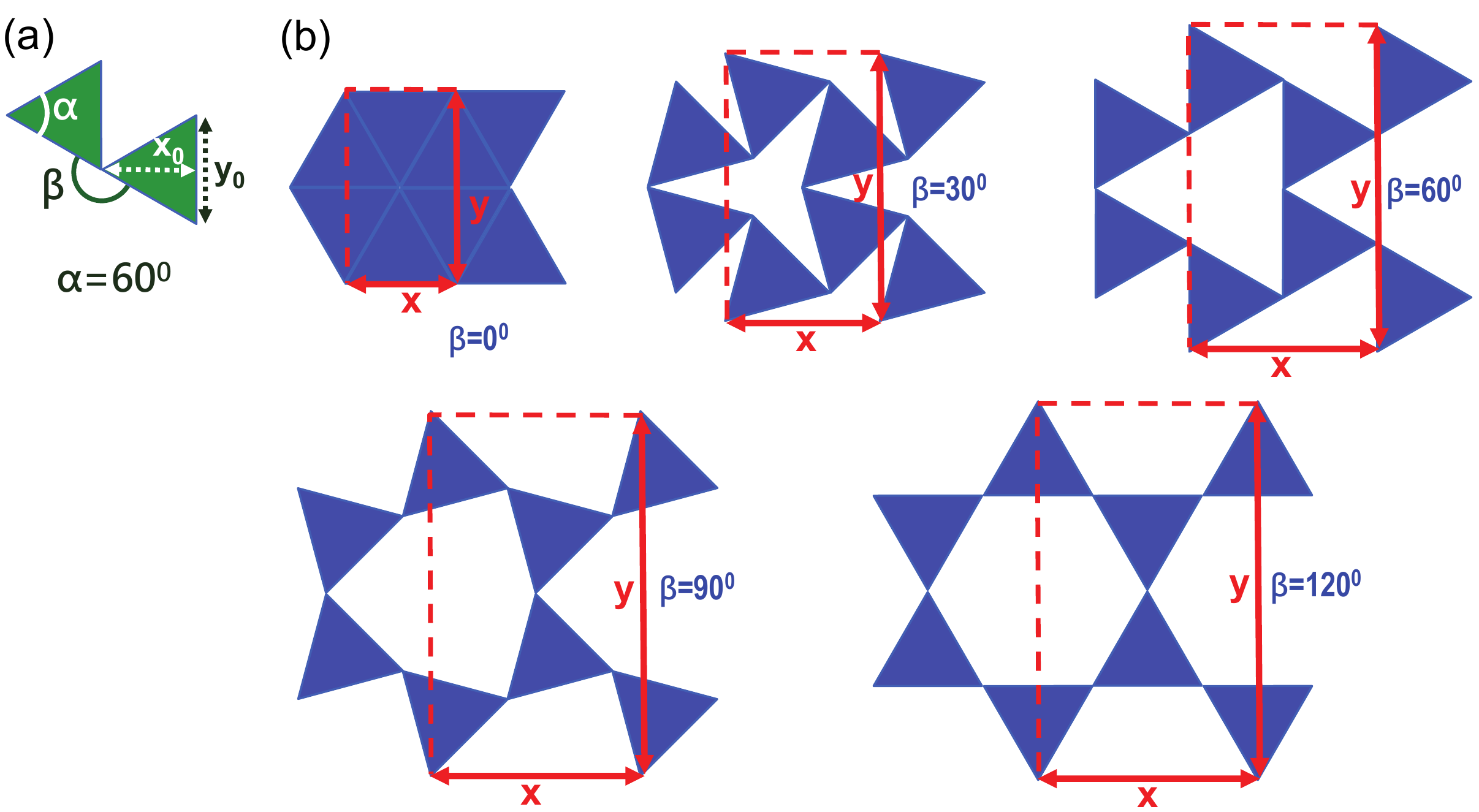}
\caption{ Deformations in a 2D assembly of rigid equilateral
triangles. %
(a) Adjacent triangles with mutual orientation defined by the
closing angle $\beta$, hinged at the corners, forming the
primitive unit cell. The triangle height $x_0$ and the length
$y_0$ of its base define the horizontal and vertical length
scales. %
(b) Snap shots of the triangle assembly for different values of
$\beta$. The conventional unit cells of width $x$ and height $y$
are indicated. %
\label{fig5} }
\end{figure}

We mentioned above that the behavior of ${\alpha}=60^\circ$
triangle systems, depicted in Fig.~\ref{fig5}, is unique among
the 2D assemblies of corner-sharing isosceles triangles. As
discussed in the main manuscript and above, the Poisson ratio
changes drastically for triangle systems with opening angle
$\alpha$ other than $60^\circ$. While hinged equilateral triangles
gradually unfold when $\beta$ increases, as seen in
Video~\ref{Video2}, the width $x$ of the unit cell remains
proportional to its height $y$, resulting in a constant,
$\beta$-independent Poisson ratio ${\nu}_{xy}=-1$, as noted
earlier~\cite{{grima2000zeolites},{Sun12369}}. For the particular
angle ${\beta}=120^\circ$, the structure of the assembly resembles
the Kagom\'{e} lattice.

\subsection{
Deformations of 2D polyphenanthrene}

Changes in the 2D polyphenanthrene structure as a function of
$\beta$ are shown in Video~\ref{Video3}. The structural changes
resemble those shown in Video~\ref{Video1} for the assembly of
${\alpha}=120^\circ$ rigid triangles.

\begin{video}[h]
\includegraphics[width=0.4\columnwidth]{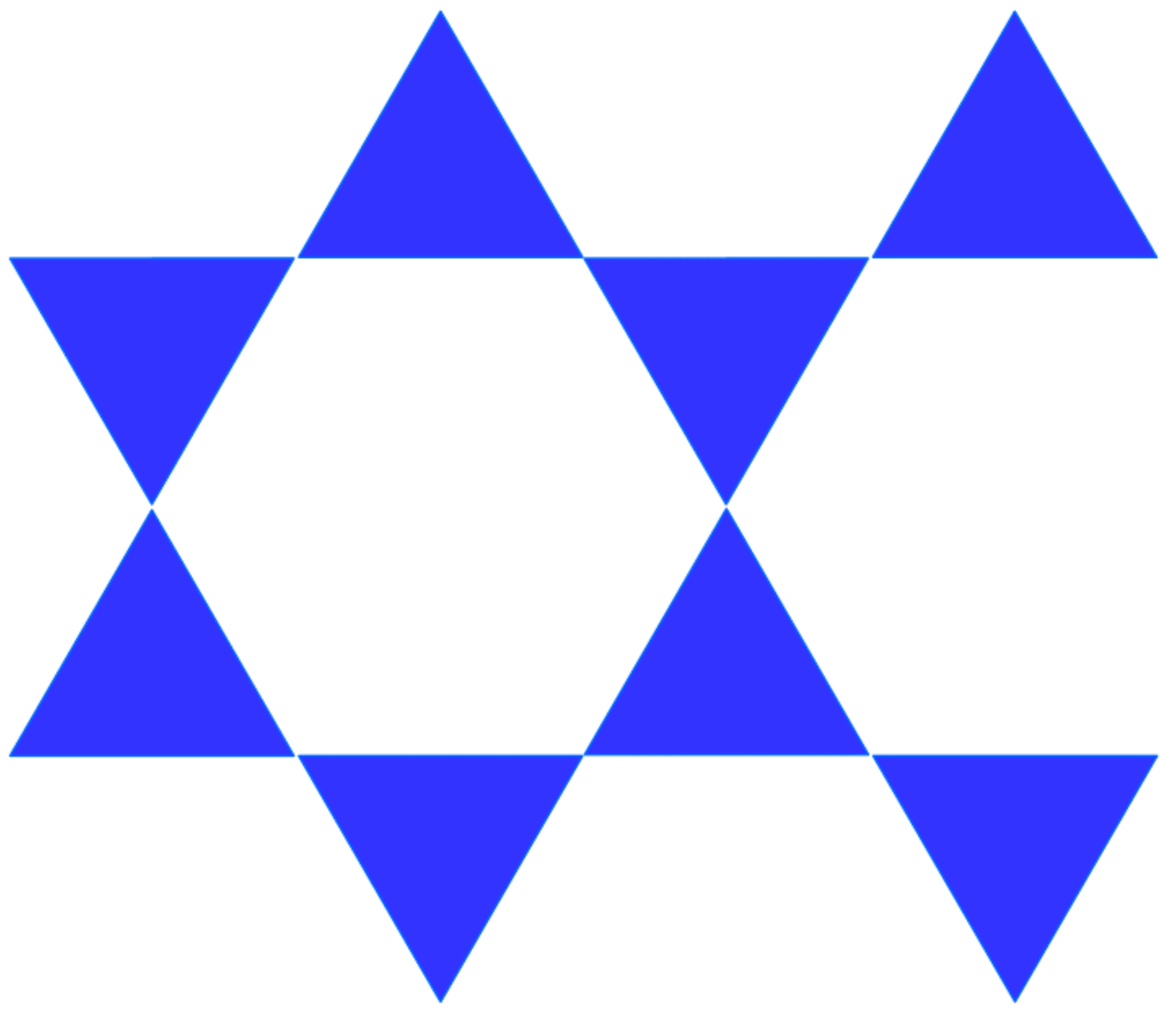}
\setfloatlink{Video2.mp4} %
\caption{Unfolding of a 2D assembly of ${\alpha}=60^\circ$
equilateral triangles with changing angle $\beta$. %
\label{Video2}}
\end{video}


\begin{video}[h]
\includegraphics[width=0.7\columnwidth]{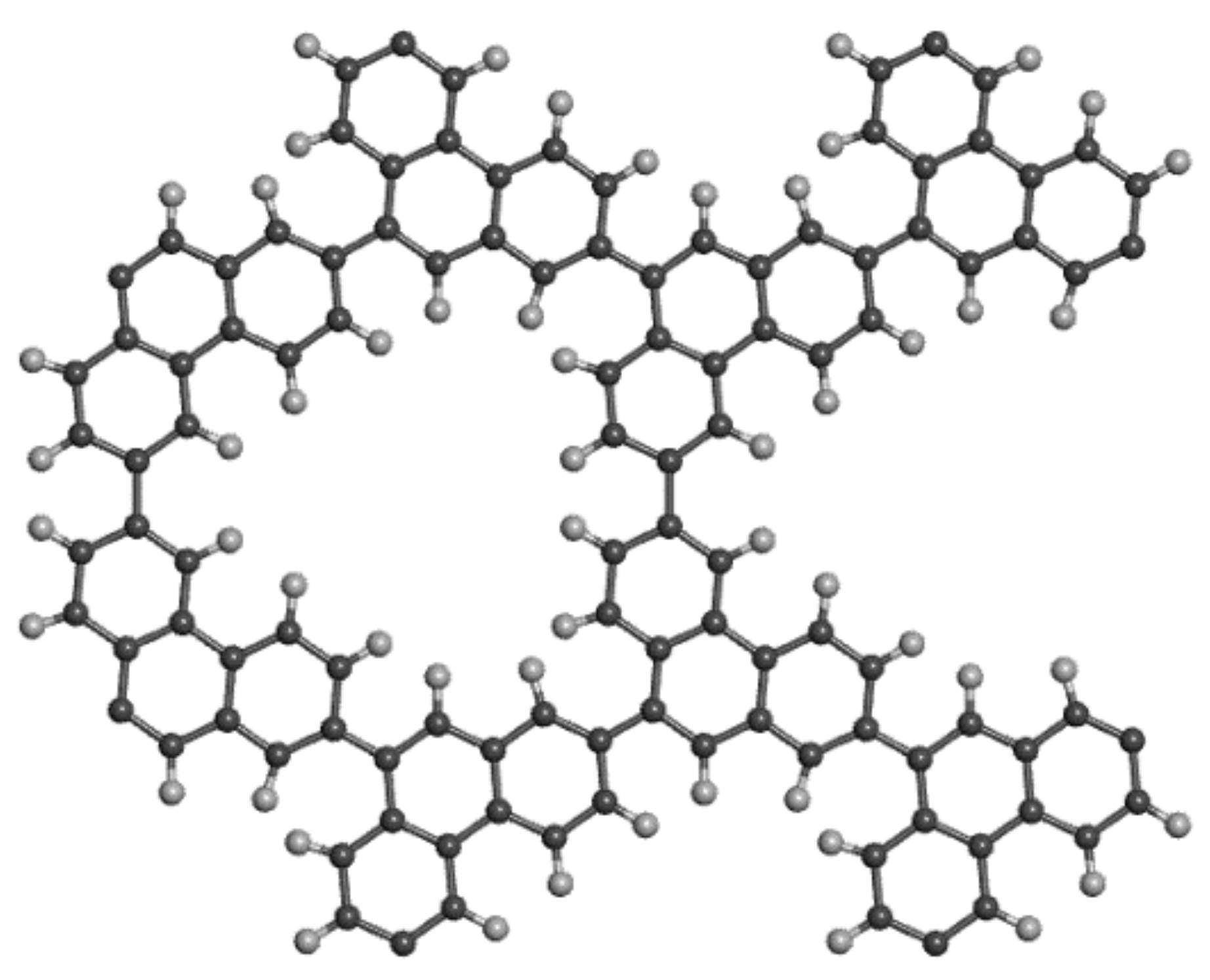}
\setfloatlink{Video3.mp4} %
\caption{Unfolding of a 2D polyphenanthrene structure dubbed
`porous graphene' with changing angle $\beta$. %
\label{Video3}}
\end{video}


\begin{acknowledgments}
We thank Jie Ren for useful discussions. D.L. and D.T. acknowledge
financial support by the NSF/AFOSR EFRI 2-DARE grant number
EFMA-1433459. Z.G. gratefully acknowledges the China Scholarship
Council (CSC) for financial support (China Scholarship number
201706260027). Computational resources have been provided by the
Michigan State University High Performance Computing Center.
\end{acknowledgments}



%

\end{document}